\documentclass{elsart}
\usepackage{amsmath}
\usepackage{amssymb}
\usepackage{amsmath}
\usepackage{verbatim}
\usepackage{epsfig}

\newcommand{\new}{\newcommand} 
 \new{\np}{\newpage} 
 \new{\md}{\medskip} 
 \new{\sm}{\smallskip}
\new{\ra}{\rightarrow} 
\new{\la}{\leftarrow}
\new{\Ra}{\Rightarrow} 
\new{\La}{\Leftarrow}
 \new{\cen}{\center}

\new{\eg}{for example}
\new{\Txs}{There exists}
\new{\txs}{there exists}
\new{\tx}{there exist} 
\new{\Tx}{There exist}
\new{\ter}{interpretation}
\new{\ters}{interpretations}
\new{\st}{such that}
 \new{\ifa}{if and only if} 
 \new{\ass}{assignment} 
 \new{\asss}{assignments}
\new{\equ}{equivalent} 
\new{\Wma}{We may assume} 
\new{\wma}{we may assume}
\new{\wrt}{with respect to}
\new{\enum}{\begin{enumerate}}
\new{\enumE}{\end{enumerate}}
\new{\sat}{satisfy}
\new{\sating}{satisfying}
\new{\otoh}{on the other hand}
\new{\Otoh}{On the other hand}
\new{\ih}{induction hypothesis}
\new{\iha}{induction hypothesis applied}

\new{\be}{\begin} 
\new{\proofend}{$\;\square$ \bigskip}
\new{\proof}{\noindent{\it Proof. }}

\new{\as}{\,{\tt {:=}}\,}
\new{\whi}{\,\mathit{while}\,}
\new{\du}{\,\mathit{do}\,}
\new{\then}{\,\mathit{then}\,}
\new{\els}{\,\mathit{else}\,}
\new{\beginn}{\text{\it begin }}
\new{\eend}{\text{\it end }}
\new{\tru}{{\mathsf T}}
\new{\fal}{{\mathsf F}}
\new{\si}{\,\mathit{if}\,}
\new{\ski}{\mathit{skip}}
\new{\sch}{\mathit{Sch}}  
\new{\Int}{\mathit{Int}}
\new{\term}{\mathit{Term}} 
\new{\sig}{\mathit{Ass}}

\new{\p}{\mathcal{P}}
\new{\f}{\mathcal{F}}
\new{\labels}{{\mathcal L}}

 \new{\var}{\mathcal{V}} 
\new{\emp}{\varepsilon}

\new{\tl}{\triangleleft}
 \new{\fA}{\Gamma}
 \new{\fS}{S}
 \new{\fT}{T}
 \new{\lang}{{\mathit alphabet}}
 \new{\allpath}{\Pi^\omega}
 
 \new{\pathset}{\Pi}
 \new{\pathinf}[1]{\Pi^{\infty}(#1)} 
 \new{\path}[2]{{\pi}_{#1} ( #2)}        
 
 \new{\set}[1]{\{ #1\}}
 
 \new{\subterms}{\mathit{Subterms}}
\new{\terms}{\mathit{Terms}}
 \new{\termfun}{\mathit{TermSymbols}}
 \new{\pathlets}{\mathit{PathSymbols}}
 \new{\predpass}{\mathit{Passpred}}
 \new{\funpass}{\mathit{Passfunc}}
 
 \new{\terlets}[1]{{\mathit{InterLetters}}_{#1}}
 \new{\sub}[1]{\,{\mathit{sub}}_{#1} \,}
 \new{\subtrans}[1]{\,{\mathit{sub}^*}(#1) \,}
 
 \new{\state}{{\rm State}}
 \newcommand{\quine}[1]{ [ \!  [{#1}] \!  ] }
 
 \new{\arity}{\mathit{arity}}
 
 \new{\schema}{\mathit{schema}}
 
 \new{\predpath}[1]{\mathit{Predpaths}_{\, #1}}
\new{\func}{\mathit{Funcs}}
 \new{\pred}{\mathit{Preds}}
 \new{\ifpred}{\mathit{ifPreds}}
 \new{\whipred}{\mathit{whilePreds}}
 \new{\dd}[2]{{\mathcal M}\quine{#1}^{#2}_d}
 \new{\ee}[2]{{\mathcal M}\quine{#1}^{#2}_e}     
  \new{\mean}[3]{{\mathcal M}{\quine{#1}^{#2}_{{#3}}}}  
 \new{\supe}[3]{#1_{#2}^{#3}} 
 \new{\pre}{\mathit{pre}}
\new{\maxpre}{\mathit{maxpre}}
 
\newcommand\vx{{\mathbf{x}}}
\newcommand\vy{{\mathbf{y}}}

\new{\vt}{{\mathbf{t}}}
\newcommand{\va}{\mathbf{a}}
\newcommand{\vb}{\mathbf{b}}

\newcommand{\vu}{\mathbf{u}}
\newcommand{\vs}{\mathbf{s}}
\newcommand{\vv}{\mathbf{v}}
\newcommand{\vw}{\mathbf{w}}
\newcommand{\vtt}{\mathbf{\tilde{t}}}

\new{\ul}{\underline}

\newcommand{\perc}[1][S]{{\underset{#1}{\rightsquigarrow}}}
\new{\Perc}[1][S]{ !!\rightsquigarrow_{#1} }
\new{\finperc}[1][S]{\rightsquigarrow_{#1}^{\text{final}}}
\newcommand{\percTwo}[2]{\underset{#1}{\overset{#2}{\rightsquigarrow}}}

\newcommand{\percFinal}[1][S]{\percTwo{#1}{\text{final}}}

\new{\cont}[1][S]{ \, \searrow_{#1} \, }
\new{\tcont}[1][S]{ \, {\searrow_{#1}}^{*} \; }

\new{\thru}[1][\fS]{\longmapsto_{#1}}
\new{\weak}[1]{\cong_{#1}}
\new{\strong}[1]{\mathbf{\cong_{#1}}}
\new{\seq}{\mathit{Sequences}}
\new{\mseq}{\mathit{maxSequences}}
\new{\isdefd}{\stackrel{\mathrm{def}}{=}}
\new{\assig}[1][S]{ \mathit{assign}_{#1} }
\new{\refset}[1][S]{ \mathit{refVars}_{#1} }
\new{\refvec}[1][S]{ \mathbf{refvec}_{#1} }

\new{\sli}[1][V]{ {{\mathit{sl}_{#1}}} }
\new{\inv}[1][S]{\mathit{Inv}_{#1}}
\new{\comps}{\mathit{Comps}}
\new{\opn}{ < \! \!\!}
\new{\clo}{\!\! >}
\new{\eqspace}{\!\! =\!\!}
\new{\simi}[1][R,V]{\tilde_{#1}}
\new{\need}[1][S]{{\mathcal N}_{{#1}}}
\new{\tword}[3]{\opn{#1}\supe{({#2})}{}{#3}\clo}
\new{\occ}{\mathit{Occ}}
\new{\gr}[1][S]{\mathit{grade}_{#1}}
\new{\grx}[1][S]{\mathit{grade}^{\mathit EXT}_{#1}}
\new{\grint}[1][S]{\mathit{grade}^{\mathit INT}_{#1}}
\new{\dex}[1][S]{\mathit{index}_{#1}}
\new{\dexx}[1][S]{\mathit{index}^{\mathit EXT}_{#1}}
\new{\inte}{\mathit{int}}
\new{\exte}{\mathit{EXT}}
\new{\seg}{\mathit{Seg}}
\new{\comb}{\bowtie}
\new{\letter}[2]{\opn {#1}\,{\clo}\!\!_{#2}\,}
\new{\pletter}[2]{\opn {#1},#2\,{\clo}}
\new{\body}[1][S]{ {{\mathit{body}_{#1}}} }
\new{\tf}{\{\tru,\fal\}}
\new{\op}{\neg}

\new{\outif}[1][S]{ {{\mathit{outif}_{#1}}} }
\new{\outwhi}[1][S]{ {{\mathit{outwhile}_{#1}}} }
\new{\out}[1][S]{ {{\mathit{out}_{#1}}} }
\new{\throu}[1][S]{ {{\mathit{thru}_{#1}}} }
\new{\back}[1][S]{ {{\mathit{back}_{#1}}} }
\new{\con}{\mathcal C}
\new{\varr}{\mathit whileVar}
\new{\corr}[1][S,T]{\mathit{Corr}_{#1}}
\new{\scorr}[1][S,T]{\mathit{StrongCorr}_{#1}}
\new{\precc}[1][S]{<\!\!<_{#1}}
\new{\abuv}[1][S]{\mathit{above}_{#1}}

\new{\tail}[1][S]{\mathit{tail}_{#1}}
\new{\head}[1][S]{\mathit{head}_{#1}}
\new{\tailterm}[1][S]{\mathit{Tailterms}_{#1}}
\new{\headterm}[1][S]{\mathit{Headterms}_{#1}}
\new{\com}[1][S]{\mathit{tail}_{#1}}

\new{\simbol}{\mathit{symbol}}
\new{\lsimbol}{\mathit{symbol}^{(\l)}}
\new{\sym}[1][S]{\mathit{Symbols}_{#1}}
\new{\lsym}[1][S]{\mathit{{Symbols}^{\mathcal L}}_{#1}}
\new{\ifprec}[1][p,S]{\sqsubseteq_{#1}}
\new{\pass}[1][S]{\mathit{passage}_{#1}}
\new{\Pass}{\mathit{Passages}}
\new{\init}{\mathit init}
\new{\next}[1][S]{{{\mathit{nextsymbol}}_{#1}}}

\new{\fir}{\mathit{firstpred}}
\new{\swhi}{\mathit{swhilePreds}}
\new{\ia}{immediately above}
\new{\sw}{super-while}

\new{\simil}[1][u]{\,\mathord{\mathit{simil}_{#1}\,}}
\new{\class}{\mathit{class}}
 \new{\trunc}[1][S]{\mathit{trunc}_{#1}\,}
 \new{\dep}[1][S]{\mathit{depnum}_{#1}\,}

 \new{\px}{p(\vx)}
 \new{\py}{p(\vy)}
 \new{\qx}{q(\vx)}
 \new{\qy}{q(\vy)}

 \new{\subb}{\subseteq}
  \new{\supp}{\supseteq}

  \new{\goto}{\mathit{goto\,\,}}

 \new{\ppair}[1][S]{\mathit{Predpairs}_{#1}}

\new{\cng}[1][u]{\mathit{cong}_{#1}\,}

\new{\simclass}[2][S]{[#2]_{#1}}

\new{\sto}{\mathit{STOP}}
\new{\start}{\mathit{START}}

\new{\before}[1][S]{<\!\!<_#1}

\new{\weis}[1][S]{{\mathcal W}_{#1}}
\new{\wpath}[1][S]{{\mathcal W}{\mathit paths}_{#1}}
\new{\wpred}[1][S]{{\mathcal W}{\mathit preds}_{#1}}
\new{\wfunc}[1][S]{{\mathcal W}{\mathit funcs}_{#1}}
\new{\wsym}[1][S]{{\mathcal W}{\mathit symbols}_{#1}}

\new{\lfnl}{LFn-L}
\new{\lib}{{\mathit libFuncs}}
\new{\cfunc}{{\mathit constFuncs}}

\def\labl{^{(l)}}
\def\labL{{^{({\mathcal L})}}}
\new{\ls}[2][l]{{#2}^{(#1)}}
\new{\labs}{\mathit{labels}}          

 \new{\lfunc}{{\func^{\mathcal L}}}
\new{\lpred}{\pred^{\mathcal L}}
 \new{\lifpred}{{\ifpred^{\mathcal L}}}
 \new{\lwhipred}{{\whipred^{\mathcal L}}}

\new{\resp}{respectively}

\new{\gra}{\mathit Graph}
\new{\simcl}[2][S]{[#2]_{#1}}

 \new{\last}{\mathit{end}}
\new{\proj}[1][S']{ {\mathit{proj}}_{#1} }

\new{\scheq}{\text{:}\!\!\!=}

\new{\ubef}{\before[\text{}]}
\new{\usclass}[1]{[#1]}

\new{\call}{{\mathit call}}
\new{\uncall}{{\mathit uncall}}

\new{\conc}{{\parallel}}

\new{\sed}[1][S]{ \exists {\mathit{DD}}_{#1}    }

\new{\sad}[1][S]{ \forall {\mathit{DD}}_{#1}    }

\new{\strt}{\mathbf{start}}
\new{\nd}{\mathbf{end}}

\new{\emptwrd}{\varepsilon}

\new{\frst}{\mathit{firstlabel}}

\new{\datdep}[1][S]{\exists\mathit{DatDep}_{#1}}
\new{\datdap}[1][S]{\forall\mathit{DatDep}_{#1}}
\new{\vrtces}{\mathit{Vertices}}

\new{\lop}{\mathbf{loop\;}}

\new{\reg}[1][{\Sigma}]{{\mathbf{Reg}} _#1           }
\new{\Reg}[1][{\Sigma}]{{\mathbf{Reg}} _#1 ^{\omega} }

\begin{document}




\begin{frontmatter}


\title{Characterizing Minimal Semantics-preserving Slices  of predicate-linear, Free, Liberal    Program  Schemas}


\thanks[acknow]{Corresponding author: Michael  Laurence, {\bf email} mike.rupen@googlemail.com, {\bf tel}+44 (0) 114 222 1800, {\bf fax} +44 (0) 114 222 1810, {\bf address}  Department of Computer Science, Regent Court, 211 Portobello, Sheffield, S1 4DP, UK.}


\author[gold]{Sebastian Danicic}
\author[bru]{Robert M Hierons}
\author[shef]{Michael R Laurence}

\address[shef]{  Department of Computer Science,
Regent Court, 211 Portobello, Sheffield, S1 4DP, UK.}


\address[gold]{              
Department of Computing,
Goldsmiths College, University of London, London SE14 6NW, UK.}

\address[bru]{Department of Information Systems and Computing, Brunel University, Uxbridge, Middlesex, UB8 3PH. }


\date{}
 
\maketitle

 \begin{abstract}
\noindent  A program schema defines a class of programs, all of which have identical statement structure, 
 but whose functions and predicates may differ. A schema thus defines an entire class of programs according to how its symbols are interpreted.  A \emph{subschema} of a schema is obtained from a schema by deleting some of its statements. 
   We  prove that given  a schema $S$  which is  predicate-linear,   free and liberal, \st\ the true and false parts of every \textit{if} predicate satisfy a simple additional condition, and a  slicing criterion defined by  the final value of a given variable after execution of any program defined by $S$,  
 the minimal subschema of $S$ which respects this slicing criterion contains all the function and predicate  symbols `needed' by the variable according to the data dependence and control dependence relations used in program slicing, which is the symbol set  given by Weiser's static slicing  algorithm. Thus this algorithm gives predicate-minimal slices for classes of programs represented by  schemas satisfying our set of conditions.  
We also give an example to  show that the corresponding result \wrt\ the slicing criterion defined by termination behaviour is incorrect. 
This complements a result by the authors in which $S$ was required to be function-linear, instead of predicate-linear.

 \end{abstract}

\begin{keyword}
program schemas
\sep
Herbrand domain
\sep
program slicing
\sep
Weiser's algorithm
\sep 
free and liberal schemas
\sep
linear schemas
\end{keyword}



\end{frontmatter}

\section{Introduction}   \label{intro.sect} 

A schema represents the 
statement structure of a program by replacing real functions and predicates by symbols representing them.  
A schema, ${S}$, thus defines a whole class  of programs which  all have the same
structure. 
Each   program  can be obtained from ${S}$ via  a domain $D$ and an {\it interpretation} $i$ which defines a function $f^i: D^n \to D$ for each function symbol $f$ of arity $n$, and a predicate function $p^i: D^m \to \tf$ for each predicate symbol $p$ of arity $m$.
As an example, Figure \ref{S.fig} gives a schema $S$, and the program $P$ of Figure \ref{P.fig} is defined from $S$ by 
interpreting the function symbols $f,g,h$ and the predicate symbol $p$ as given by $P$, with $D$ being the set of integers. 

 The subject of schema theory is connected with that of program  transformation and  was originally motivated by the wish to compile programs     effectively\cite{greibach:theory}.  Schema theory is also  relevant to program slicing. 
Since  program slicing algorithms do not normally take into account the meanings of the functions and predicates of a program, a schema encodes all the information about any program which it defines that is available to such  algorithms.

\begin{figure}[t]
\begin{center}
$
\begin{array}{llll}
u\as h(); \\
\si  p(w) && \then 
& v\as f(u); \\
&& \els 
&  v\as g();
 \end{array}
 $ 
\end{center}
\caption{Schema $S$} \label{S.fig}
\end{figure}

\bigskip


A \emph{subschema}  of a schema $S$  is defined to be any schema obtained   by deleting statements from $S$. 
Given a  schema $S$  and  a variable $v$, we wish to find a  subschema $T$ of $S$ which satisfies the following condition; given  any \ter\ and any initial state \st\   the program defined by $S$ terminates,  that defined by $T$ also does, and defines the same final value for $v$. 
 In this case we say that $T$ is a $v$-slice of $S$. We are particularly interested in finding \emph{minimal} $v$-slices of $S$ (with slices of $S$ ordered according to their sets of symbols\footnote{A {\it symbol} in this paper means a function or predicate symbol in a schema.}). 

The main theorem  of this paper  requires that given any  path through a schema $S$, there is an \ter\ and an initial state \st\ the program thus defined  follows this path when executed (the freeness condition) and 
the same term is not generated more than once as it does so (the liberality  condition). These conditions were first defined by Paterson \cite{paterson:thesis}.  We also require that  the same predicate symbol does not  occur more than once in $S$ (the predicate-linearity condition), and that if the same function symbol occurs in both the true and false parts of any \textit{if} predicate\footnote{If the statement  $\si p(v) \then T_1 \els T_2$ occurs in a predicate-linear schema $S$, then we say that $T_1$ and $T_2$ are respectively the true and false parts of $p$ in $S$}, then it assigns to distinct variables in each case. We call schemas satisfying all these conditions \emph{special} schemas.
 We prove that given a  schema $S$ which satisfies these conditions and  a variable $v$, the $v$-slice of $S$ given by   Weiser's static slicing  algorithm\cite{weiser:slicing79} has the unique  minimal set of predicate and function symbols of all $v$-slices of $S$. Given a schema, Weiser's algorithm computes the  subschema  containing only those symbols defined by the transitive closure of the control and backward data dependence relations.  We also define an $\omega$-slice of a schema in which  termination behaviour defines the  slicing criterion, and give an example to  show that  Weiser's algorithm,  modified in a natural way \wrt\ this  slicing criterion, need  not give a minimal $\omega$-slice. This is in contrast to the situation for function-linear, free, liberal schemas \cite{laurence:flfl}. 

\begin{figure}[t]
\begin{center}
$
\begin{array}{llll}
& \\
u\as 1; \\
\si  w> 1 && \then 
& v\as u+1; \\
&& \els 
&  v\as 2;
 \end{array}
 $
\end{center}
\caption{Program $P$} \label{P.fig}
\end{figure}

 Our theorem  is a strengthening of the result in \cite{sdetal:lpr} in which no symbol was allowed to occur more than once in the schema $S$ (that is, $S$ had to be linear, 
as opposed to just predicate-linear in this paper).  


\subsection{Organisation of the paper}

In the remainder of this section, we explain how the field of program slicing provides motivation for our results, and we also discuss the history of the study of schemas. In Section \ref{basic.sect}, we give formally our basic schema  definitions. In Section \ref{free.lib.sect} we define formally free and liberal schemas, and also give a simple characterisation of schemas that are both free and liberal, which shows that Weiser's algorithm preserves the property of being  both free and liberal for slices. In Section \ref{equiv.defn.sect} we formally define a subschema of a schema. In Section  \ref{data.defn.sect} we formally define the data dependence relations $\perc$ and $\percFinal$ for a schema $S$ and define Weiser's  labelled symbol set for a schema. We also give examples of  cases in which the subschema of a schema containing only the symbols in Weiser's set is not the minimal subschema satisfying the required conditions. In Section \ref{couple.section}, we define the notion of a $p$-couple for a predicate $p$; that is, a pair of \ters\ which differ only  on one $p$-predicate term. In Section \ref{special.sect} we introduce formally the class of special schemas to which our results apply. In Section \ref{main.sect}, we prove our main theorems. In Section \ref{omega.sect}, we give an example to  show that the subschema of a special schema given by Weiser's algorithm \wrt\ termination need not be minimal of all subschemas preserving termination behaviour. In Section \ref{conclu.sect}, we discuss our conclusions.

\subsection{Relevance of Schema Theory to Program Slicing} \label{relevance.slicing.subsect}

The field of (static) program slicing is largely concerned with the design of algorithms which given a program and a variable $v$, eliminate as much code as possible from the program, \st\ the subprogram  consisting of the remaining code, when executed from the same initial state,  will still give the same  final value for $v$ as the original program, and preserve termination.  One algorithm is thus better than another if it constructs a smaller slice.


\begin{figure}[t]
\begin{center}
$
\begin{array}{llll}
\; v\as g();\\
\si  p(u) & \then v\as g();\\
 \end{array}
 $
\end{center}
\caption{Deleting the \textit{if} statement gives a $v$-slice of this schema} \label{not.linear.fig}
\end{figure}

Most program slicing algorithms are based on the  \emph{program dependence graph} (PDG) of a program. This includes Weiser's algorithm\cite{weiser:slicing79}, which was, however, expressed in different language.  (For a fuller  discussion of program slicing algorithms see \cite{tip:slice-survey,binkley:annals}.) The PDG of a program is a graph whose vertices are the labelled statements of the program and whose directed edges indicate \textit{control} or \emph{data dependence} of one statement upon another. 

Data dependence is defined as follows. 
We say that in a schema $S$,  a function or predicate symbol  $x$ is data dependent upon a function symbol $f$, written $f \perc x$, if $x$ references the  variable to which $f$ assigns, and there is a path through $S$ passing  through $f$ before passing  through $x$   without passing through an intermediate assignment to the same variable as $f$. The relation statement $f \percFinal v$ is defined analogously for a function symbol $f$ and variable $v$ using terminal path-segments.  This definition of the relations  $\perc,\,\percFinal$  is purely syntactic;  feasability of any  path is not required for it to hold. 
Thus $h \perc f$, $\, f\percFinal v$ and $g\percFinal v$ hold for the schema of Figure \ref{S.fig}; $g\percFinal v$ means that there is a path through $S$ passing through $g$, and not subsequently passing through a later \ass\ to the variable $v$ before reaching the end of the schema.  

Slicing algorithms  do not take account of the meanings of the functions and predicates occurring in a program, nor do they exploit the knowledge that the same function or predicate  occurs in two different places in a program. This reflects the fact that it is undecidable whether the deletion of a particular line of code from a \emph{program} can affect the final value of a given  variable after execution\cite{danicic:phd}. 
\Otoh\ a 
 schema likewise encapsulates the data  and control dependence relations of the programs that it represents, but whereas it also  does not encode the meanings of its function and predicate symbols, it does record any  multiple occurrences of these symbols, and this extra  information may  sometimes lead to a proof of the existence of smaller slices.
 As an example,  it is obvious that the predicate symbol $p$ and the \ass\ that it controls may be deleted from the schema of Figure \ref{not.linear.fig} without preventing termination  or changing the final value of $v$ (that is, the resulting subschema is  a $v$-slice in our terminology), but most program slicing algorithms will treat the two occurrences of $g$ as if they were two distinct functions, and therefore will not make any deletion.

 However, slicing algorithms taking   linear schemas as input
 may yield more information about a program  than algorithms that merely use Weiser's algorithm. 
As an example, in the schema $S$ of Figure 
\ref{not.lib.fig}, which will be discussed in  further  sections, 
it can be seen that the subschema of $S$ obtained by deleting the \ass\ with symbol $f$ is a $v$-slice of $S$, since the removal of this \ass\ cannot prevent termination (which is determined solely by the value of $w$ when referenced by $q$), nor can it prevent the path of execution from passing through $g_1$ at least once, though it may affect the number of times this happens. However, 
 if the \ass\ with symbol $g_1$ is replaced by an \ass\ $v\as g_2(v);$ to give a schema $T$,  then the  \ass\ $u \as f(u);$ may not similarly  be deleted from $T$, since this deletion may change the value of $v$ after execution. As an example of an \ter\ under which this occurs, suppose that $h_1,h_2,f$ and $g_2$ are all interpreted as the function $v \mapsto v+1$ in the domain of integers and $q(0), q(1), p(0),p(1) $ and $p(2)$ map to \emph{true}, whereas $q(v)$ and $p(v)$ map to \emph{false} if $v \ge 2$ or $v \ge 3 $ respectively. Execution of $S$ from the initial state in which all variables are set to zero results in a final value of $1$ for $v$, whereas if the \ass\ $u \as f(u);$ is deleted, then the execution path will pass through $g_2$ on both occasions that it enters the body of $q$ giving a final value of $3$ for $v$. 
However Weiser's algorithm will treat these two cases identically, and will require $f$ to be in a $v$-slice in both cases. This is because for $S$ and $T$,  Weiser's set \wrt\ the variable $v$ must contain $g_1$ or $g_2$ respectively, since $g_1 \percFinal v$ and  $g_2 \percFinal[T] v$, and $p$ controls $g_1$ or $g_2$ respectively, hence Weiser's set must contain $p$, and $f \perc h_2 \perc p$ (and similarly for $T$), thus Weiser's set contains $f$. 
 Danicic \cite{danicic:phd} gives other examples of cases of linear schemas for which program slicing algorithms will not give minimal correct subschemas. If the linearity assumption is discarded, then non-minimality can be demonstrated even for loop-free schemas, such as the one in Figure \ref{not.linear.fig}, in which $p$ and both occurrences of $g$ lie in the Weiser symbol set defined by $v$, but the statement containing $p$ can clearly be deleted without changing the final value of $v$. 
 These examples motivate the mathematical study of schemas, which may lead to the computation of smaller subschemas than conventional program slicing techniques can achieve.

\begin{figure}[t]

\begin{center}
$
\begin{array}{llll}
\whi q(w)  \du  & \{ \\
& w \as h_1(w);  \\
& u \as h_2(u);  \\
&\si  p(u)  \then &\{ \\

&&v\as g_1(); \\
&&u\as f(u); \\
&&\}\\
&\}
 \end{array}
 $
\end{center}
\caption{Deleting the \ass\ $u\as f(u);$ gives a $v$-slice of this linear schema, although $u\as f(u);$ lies in Weiser's statement set \wrt\ $v$} \label{not.lib.fig}
\end{figure}


\subsection{Different classes of  schemas}   \label{diff.classes.schemas.sub}

Many subclasses of schemas have been defined:

\begin{description}
\item[Structured schemas,] in which \emph{goto} statements are forbidden, and thus loops must be constructed using \textit{while} statements. \emph{All schemas considered in this paper are structured.} 
 \item[Linear schemas,]  in  which   each 
function and predicate symbol occurs at most once.
\item[Predicate-linear schemas,] which we introduce in this paper,   in  which   each 
predicate symbol  occurs at most once, but which may have more than one occurrence of the same  function symbol.
\item[Free schemas,]  where all paths are executable under some interpretation.
\item[Liberal schemas,] in which two \asss\ along any legal path can always  be made to assign distinct values to their respective variables.
\item[Near-liberal schemas,] which the authors introduced in \cite{laurence:danicic:hierons:nlfl:jlap},   in which this non-repeating condition applies only to terms not having the form $g()$ for a  function symbol $g$ of zero arity.
\end{description}

We now give examples of schemas satisfying these definitions, and first show that 
the freeness and liberality conditions on schemas are incomparable. To see this, consider the following two examples of linear schemas. The schema $$\whi p(v) \du \ski$$
 contains no \asss\ and is therefore liberal,  but it is not free, since there is no choice of \ter\ and initial state under which the executed path thus defined passes exactly once through the body of $p$, since the value of $v$, and hence the boolean value defined at $p$ cannot change during execution. 
\Otoh\ the schema  
\begin{center}
$
\begin{array}{llll}
\whi q(w) \du & \{  \\
 &w\as f(w);\\
&x\as g();\\
&\}
 \end{array}
 $
\end{center} is free, since if $f$ defines the function $w \mapsto w+1$ over the domain of integers, then $w$ never defines a repeated value when referenced by $q$, and so $q$ can be interpreted so as to define an executed path that passes any desired number of times through $q$, 
 but it is not liberal, since the variable $x$ is always assigned the same value at occurrences of $g$ along any executed path. 
The subschema obtained from it by deleting the \ass\ $x\as g();$ (that is, $\whi q(w) \du w\as f(w);$) is both free and liberal, \otoh.\

 The schema in Figure \ref{not.lib.fig} can also be seen to be free, owing to the self-referencing \asss\ with symbols $f,h_1,h_2$, which can be interpreted as the function $w \mapsto w+1$ over the domain of integers, thus ensuring that the variables $u,w$ referenced by $p$ and $q$ respectively never repeat in value. It is not liberal however, since it has a path passing more than once through $g_1$, along which this \ass\ defines the same value to $v$ on each occasion. More generally, it is easy to see that no schema having a constant \ass\ in the body of a \textit{while} predicate can be both free and liberal, since if it is free, then there is an executable path passing twice through this \ass,\ which clearly assigns the same value to its variable on each occasion.

Two schemas are said to be \emph{equivalent} if they have the same termination behaviour, and give the same final value for every variable, given every symbol \ter\  and initial state.  The authors have shown \cite{mletal:cfl,lfl:tr} that it is decidable whether  linear, free, liberal  schemas are equivalent.

Paterson \cite{paterson:thesis} 
gave a proof that it is decidable  whether a   schema is both liberal and  free (which we give in Section \ref{free.lib.sect}); and since he also gave an algorithm transforming a schema $S$ into  a schema $T$ \st\ $T$ is both liberal and  free \ifa\ $S$ is  liberal, it is clearly decidable whether a schema is liberal. It is an open problem whether   freeness is decidable for the class of  linear   schemas. However he also proved, using a reduction from the Post Correspondence Problem,  that it is not decidable whether an arbitrary schema is free.


\subsection{Previous results on the decidability of  schema equivalence}

Most previous research on schemas has focused on schema equivalence, as defined in Section \ref{diff.classes.schemas.sub}. 
All  results on the decidability of equivalence of schemas are either negative or confined to very restrictive classes of schemas. In particular   
Paterson \cite{paterson:thesis} proved  that  equivalence 
is undecidable for the class of all  (unstructured)   schemas.  He proved this by showing that the halting problem for Turing machines (which is, of course, undecidable) is reducible to the equivalence problem for the class of all  schemas. 
Ashcroft and Manna showed  \cite{ashcroft:while-goto-siam} that an arbitrary schema can be effectively transformed into an equivalent structured schema, provided that statements such as  $\whi \neg  p(\vu) \du T$ are permitted; hence Paterson's result shows that any class of schemas for which equivalence can be decided must not contain this class of   schemas.  Thus in order to get positive results on this problem, it is plainly necessary to define the relevant  classes of schema with great care. 
\\
Positive results on the decidability of equivalence of schemas include the following; 
 in an early result in schema theory, Ianov \cite{ianov:logical} introduced a 
restrictive class of schemas, the Ianov schemas,  for which equivalence
is decidable. This problem was later shown to be NP-complete \cite{rutledge:ianov.schemata,constable:ianov.schemas}.

 Paterson \cite{paterson:thesis} proved that equivalence is decidable
for a class of schemas called {\em progressive schemas}, in which every assignment
references the variable assigned by the previous assignment
along every legal path. 

Sabelfeld \cite{sabelfeld:algorithm}  
 proved that equivalence is
decidable for another class of schemas called {\em through
schemas}. A  through schema satisfies two conditions: firstly, 
that on every path from an accessible predicate $p$ to a predicate $q$
which does not pass through another predicate, 
and every variable $x$ referenced by $p$, there is a variable referenced by $q$ which defines a term containing the term defined by 
 $x,$ and secondly, 
 distinct variables referenced by a 
 predicate can be made to  define distinct terms under some \ter.
 
 In view of the evident difficulty of obtaining positive results on this problem, and the importance of program slicing, it seems sensible to concentrate on trying to decide equivalence for classes of schema pairs in which one schema is a subschema of the other, as was done for a class of near-liberal schemas in \cite{laurence:danicic:hierons:nlfl:jlap}.

 \section{Basic  definitions  for schemas} \label{basic.sect}

Throughout this paper,  $\f$, $\p$,  $\var$ and $\labels$ denote 
fixed infinite sets of \emph{function symbols},  
\emph{predicate symbols}, \emph{variables} and \emph{labels} respectively. 
We assume  a  function
$$\arity: \f \cup\p \to \Bbb{N}.$$
The arity of a symbol $x$ is the number of arguments referenced 
by $x$.
 Note that in the case when the arity of a function symbol $g$ is zero,  
 $g$ may be thought of as a constant. 

 The set $ \term({\f}, {\var})$ of \emph{terms} is defined as follows: 
\begin{itemize}
\item each variable is a term, 
\item
if $ f\in {\f}$ is of arity $
n$ and $ t_1,\ldots ,t_n$ are terms then $ f(t_1,\ldots ,t_n)$ is a term. 
\end{itemize}

We refer to a tuple $\vt =(t_1,\ldots, t_n)$, where each $t_i$ is a term, as a vector term. We call $p(\vt)$ a predicate term if $p\in \p$
and the number of components of the  vector term $\vt$  is $\arity(p)$. 

We also  define $F$-terms and $vF$-terms recursively for $F\in \f^* $ and $v\in \var$. Any term $f(t_1,\ldots ,t_n)$ is an $f$-term, and the term $v$ is a $v$-term. 
If $g\in \f$ and  at least one of the terms $t_1,\ldots ,t_n$ is an $F$-term or $vF$-term,  then the term $g(t_1,\ldots ,t_n)$ is an $Fg$-term, or $vFg$-term, respectively.  
Thus any $FF'$-term is also an $F'$-term.

 \begin{defn} [schemas]\label{struc.sch.def} \rm
We define the set  of all \emph{schemas} recursively as follows. 
$\ski$ is a schema. 
An   assignment $y\as \ls{f}(\vx);$ where $y\in\var$, $f\in \f$, $l\in \labels$  and $\vx$ is a vector of $\arity(f)$ variables,  is a schema. 
From these all schemas 
may be `built up' from the 
following constructs on schemas. 

\begin{description}
\item[sequences;] 
$S'=U_1 U_2\ldots U_r $ is a schema  provided that 
 each $U_i$ for $i\in \{1,\ldots, r\}$ is a  schema.  
\item[{\it if} schemas;] $S'' =\si  \ls{p}(\vx) \then \{T_1\} \els \{T_2\}$ is a schema 
 whenever $p\in \p$, $l\in \labels$, $\vx$ is a vector of $\arity(p)$ variables, 
and $T_1,T_2$ are schemas. We call the schemas $T_1$ and $T_2$  the \emph{true} and \emph{false} parts  of  $p\labl$. 
\item[{\it while} schemas;] $S'''= \whi \ls{q}(\vy) \du \{T\}$ is a schema  
whenever $q\in \p$, $l\in \labels$, $\vy$ is a vector of $\arity(q)$ variables,
and $T$ is a schema. We call $T$ the \emph{body} of the \emph{while} predicate  $q\labl$ in $S'''$. If $x$ is a labelled symbol in $T$, and there is no labelled \textit{while} predicate $\ls[m]{p}$ in $T$ which also contains $x$ in its body, then we say that $\ls{q}$ lies \emph{immediately} above $x$. 
\end{description}
Thus a schema is a word in a language over an infinite alphabet. 
We normally omit the braces $\{$ and $\}$ if this causes no ambiguity.   Also, we may write    $\si  p^{(l)}(\vx) \then \{T_1\} $ instead of   $\si  p^{(l))}(\vx) \then \{T_1\} \els \{T_2\}$ if $T_2 = \ski$. 
\end{defn} 

If no symbol   
  (that is, no element of $ \f \cup\p$)  appears 
more than once in a schema  $ S$, then $S$ is said to be \emph{linear}. If  no element 
of $ \p $  appears
more than once in a schema  $ S$,  then $S$ is said to be \emph{predicate-linear}. We define \emph{function-linear} schemas analogously using the set $\f$.

The labels on function and predicate symbols do not affect the semantics of a schema; they are merely included in order to distinguish different occurrences of the same  symbol in a schema; {\it we always assume that distinct occurrences of a  symbol in a schema have distinct labels}. We will often omit labels on symbols in contexts where they need not be referred to,  as in Figure \ref{not.linear.fig}, or where a symbol only occurs once in a schema. 
In particular, our main theorems assume predicate-linear schemas, hence we do not label predicate symbols in Section \ref{main.sect}.

We define    $\sym[](S)=\func(S)\cup\pred(S)$,  $\func(S)$ and   $\pred(S)$ to be the sets of symbols,  function symbols and predicate  symbols occurring in a schema $S$.  Their labelled counterparts are  $\lsym[](S)$,  $\lfunc(S)$ and   
$\lpred(S)$. Also $\lifpred(S)$ and   $\lwhipred(S)$ are the sets of all labelled \textit{if} predicates and \textit{while} predicates in $S$.  A schema without predicates (that is, a schema which consists of a sequence of \asss\ and $\ski$s)  is called \emph{predicate-free}.

If a  schema $S$ contains an \ass\ $y\as f\labl(\vx);$ then we define $y=\assig(\ls{f})$ and 
$\vx = \refvec(\ls{f}) $. 
 If $p\labl \in \lpred(S)$ then $\refvec(p\labl)$ is  defined similarly.



 \begin{defn}[the $\cont$  relation] \rm  \label{cont.def} \mbox{}\\
 Let $S$ be a  schema.
If $\ls{p} $ is a labelled predicate in $S$ and $x$ is any (possibly labelled) symbol, we say that  $\ls{p} \cont x$ holds   if $x$ lies in the body of $\ls{p}$ (if $\ls{p}$ is a \textit{while} predicate in $S$) or $x$ lies in the true or false part of $\ls{p}$ (if $\ls{p}$ is an \textit{if} predicate). We may strengthen this by writing 
 $\ls{p} \cont x\, (Z)$ for $Z\in \tf$ to indicate   the additional condition that $x$ lies in the $Z$-part of $\ls{p}$ if $\ls{p}\in \lifpred(S)$, or  $\ls{p}\in \lwhipred(S)$ (if $Z=\tru$). 
\end{defn}

The   relation $ \cont$ is the transitive closure of the relation `controls' in program analysis terminology. 

 \subsection{Paths through a Schema}

The execution of a program defines a possibly infinite sequence of 
assignments and predicates. Each such sequence will correspond to a \emph{path}  
through the associated schema. The set $\allpath(S)$ of paths through $S$ is now given.

 \begin{defn}[the set $\allpath(S)$ of paths through   $S$, path-segments of $S$]
  \rm \label{path.defn}  
 If $L$ is any  set, then we write $L^*$ for the set of  finite words over $L$ and  $L^\omega$ for 
the set containing both finite and   infinite words over $L$.
 If $\sigma$ is a word, or a set of words over an alphabet, 
 then $\pre(\sigma)$ is the set of all finite prefixes of
(elements of) $\sigma$.

 For each schema $S$ the alphabet of $S$, written $\lang(S)$ is the set 
$$\{ \ul{y \as \ls{f}(\vx)} \vert~ y\as \ls{f}(\vx); \text{ is an assignment in }
  S\}$$ $$ \bigcup $$ $$\{\ul{p^{(l)},Z}\;  \vert~  \ls{p}\in \lpred(S)\wedge Z\in \tf \}.$$
We define  $\simbol(\ul{y \as \ls{f}(\vx)})=f$ and $\simbol(\ul{p^{(l)},Z})= p$. 

The words in $\pathset(S) \subseteq (\lang(S))^*$ 
are formed by concatenation from the 
words of subschemas of $S$ as follows: 

\begin{description}
\item[For $\ski$,] \[\pathset(\ski)\] is the set containing only the empty word. 
\item[For assignments,] \[\pathset(y\as \ls{f}(\vx);)=\{ \ul{y\as \ls{f}(\vx)}\}.\] \\
\item[For sequences,] 
$\pathset(S_1S_2\ldots S_r)=\pathset(S_1)\ldots \pathset(S_r)$. \\
\item[For \emph{if} schemas,] $\pathset(\si  p\labl(\vx) \then   \{T_1\} \els \{T_2\})$ is the 
set of all concatenations of $\ul{p\labl,\tru}$ with a word in 
$\pathset(T_1)$ and all concatenations of $\ul{p\labl,\fal}$
with a word in $\pathset(T_2)$. \\
\item[For \emph{while} schemas,]\ $\pathset(\whi q\labl(\vy) \du \{T\})=(\ul{q\labl,\tru }\, \pathset(T))^*\ul{q\labl,\fal }$. 
\end{description}
We define 
$\allpath(S)= \{\sigma \in (\lang(S))^\omega \vert \pre(\sigma) \subseteq \pre(\pathset(S))\}$.
Elements of $\allpath(S)$ are called 
{\it paths} through  $ S$. Any  $\mu\in \lang(S)^*$ is a \emph{path-segment} (in $S$) if there are words $\mu',\mu''$ \st\ $\mu'\mu\mu''\in \pathset(S)$. A \emph{terminal} path-segment of $S$ is a path-segment $\nu$ \st\ $\mu \nu\in  \pathset(S)$ for some $\mu$. 
\end{defn}

 \subsection{Semantics of   schemas} \label{struc.sch.sem.subsec}

The symbols upon which schemas are built are given meaning by 
defining the notions of a state and of an interpretation. It 
will be assumed that `values' are given in a single set $D$, 
which will be called the \emph{domain}. We are mainly interested in the case in which $D=\term(\f,\var)$ (the Herbrand domain) and the function symbols represent the `natural' functions \wrt\ $\term(\f,\var)$.

 \begin{defn}[states, (Herbrand) \ters\ and the natural state $e$] \rm 
 \label{state.ter.defn} 
\mbox{}\\
Given a domain $D$, a \emph{state} is either $\bot$ (denoting non-termination) or a function $\var \ra D$. The set 
of all such states 
will be denoted by $\state(\var,D)$. 
An interpretation $i$ defines, for each function symbol 
$f\in\f$ of arity $n$, a function $ f^i:D^n \ra D$, and for each 
predicate symbol $p\in\p$ of arity $m$, a function 
$ p^i:D^m \ra \{\tru ,\, \fal \}$. The set of all 
interpretations with domain $D$ will be denoted 
$\Int({\f}, {\p}, D)$. 
\\ 
We call the set $\term(\f,\var)$ of terms the \emph{Herbrand domain}, and we say that a function from $\var$ to $\term(\f,\var)$ is a Herbrand state. 
An interpretation $i$ for the Herbrand domain is said to be \emph{Herbrand}   if the functions $f^i:\, \term(\f,\var)^n \to \term(\f,\var)$ for each $f\in \f$
are defined as 
\begin{center}$f^i(t_1,\ldots,t_n)= f(t_1,\ldots,t_n)$ \end{center}
for all $n$-tuples of terms $(t_1,\ldots,t_n)$.
\\
We define 
the \emph{natural state} $e:\var \ra \term(\f,\var)$ 
 by $e(v)=v$ for all $ v\in \var.$ 
\end{defn}

Note that an \ter\ $i$ being  Herbrand  places no restriction on the mappings \\
$ p^i:(\term({\f}, \var))^m \ra \{\tru ,\, \fal \}$ defined by   $ i$ for each $p\in \p$.

Given a schema $S$ and a domain $D$, an 
initial state $d\in\state(\var,D)$ with $d\not= \bot$ and an interpretation 
$i\in\Int({\f}, {\p}, D)$ we now define the final state 
$\dd{S}{i}\in\state(\var,D)$ and the associated   path 
$\path{S}{i,d}\in\allpath(S)$. In order to do this, we need to define the predicate-free schema associated with the prefix of a path by considering the sequence of \asss\ and $\ski$s through which it passes.

\begin{defn}[the schema $\schema(\sigma)$]\rm \mbox{}\\ \label{schema.of.path.defn}
Given a word $ \sigma \in (\lang(S))^*$ for a schema $S$, we recursively define  the predicate-free 
schema $\schema(\sigma)$ by the following rules; $\schema(\lambda)= \ski$ if $\lambda$ is the empty word, 
 $\schema(\sigma \ul{ v\as {f}(\vx)}) \;= \; \schema(\sigma)\, v\as {f}(\vx);$ and \\
$\schema(\sigma \ul{p\labl,X})\; = \; \schema(\sigma)$. 
\end{defn} 

\begin{lem} 
\label{exe.lem}
Let $S$ be a schema.
 If  $\sigma \in \pre(\pathset(S))$,  the set 
$\{ m\in \lang(S) \vert\, \sigma m \in \pre(\pathset(S)) \}$ is 
one of the following;  a singleton containing  an underlined assignment,  a  
 pair  $\{ \ul{\ls{p},\tru} ,\;\ul{\ls{p},\fal} \}$ where $\ls{p}\in \pred\labL(S)$, or the empty set, and if $\sigma \in \pathset(S)$ then the last case holds. 
\end{lem}

Lemma \ref{exe.lem}, which was proved in \cite[Lemma 6]{laurence:flfl}, 
reflects the fact that at any point in the execution of a program, 
there is never more than one `next step' which may be taken, and an element of $\pathset(S)$ cannot be a strict prefix of another.


 \begin{defn} [semantics of predicate-free schemas] 
 \rm \label{meaning} 
Given a state $d\not= \bot$,  the final state $\dd{S}{i}$ and associated path $\path{S}{i,d} \in \allpath(S)$ 
of a schema $S$ are defined 
as follows:\\
\begin{description}
\item{For $\ski$,} $$\dd{\ski}{i}= d$$
		\begin{center}and\end{center} $$\path{\ski}{i,d} \mbox{ is the empty word.}$$

\item{For assignments,}
$$ \dd{y \as \ls{f}(\vx);}{i} (v)~~~=~~~
\begin{cases} d(v) & \text{if $ v \not= y$}, 
\\  f^{i}(d(\vx))  & \text{if $ v=y$ } 
 \end{cases}$$
 
 (where the vector term $d(\vx)=(d(x_1),\ldots, d(x_n))$ for $\vx=(x_1,\ldots, x_n)$)
\begin{center}and\end{center}
\[\path{y\as \ls{f}(\vx);}{i,d}~~~=~~~  y\as \ls{f}(\vx),\]
\item
{and for sequences $S_1S_2$  of predicate-free schemas,}  
\[\dd{S_1S_2}{i}~~~=~~~{\mathcal M}\quine{S_2}^{i}_{\dd{S_1}{i}}\]
\begin{center}and\end{center}
\[\path{S_1S_2}{i,d}~~~=~~~\path{S_1}{i,d}\path{S_2}{i,\dd{S_1}{i}}.\] 
\end{description}
This uniquely defines $\dd{S}{i}$ and $\path{S}{i,d}$ if $S$ is predicate-free.
\end{defn}

 In order to give the semantics of a general schema  $S$,
 first the path, $ \path{S}{i,d}$, of $S$ 
 with respect to interpretation, $i$, and 
 initial state $d$ is defined.

\begin{defn}[the path $ \path{S}{i,d}$]
\rm
Given  a schema $S$, an interpretation $i$, and a 
state, $d\not= \bot$, the path
 $  \path{S}{i,d} \in  \allpath(S)$ is defined by the   following condition;  for all  $\sigma \, \ul{\ls{p},X} \; \in \pre(\path{S}{i,d})$, the equality  $p^i(\dd{\schema(\sigma)}{i}(\refvec(\ls{p})))=X$ holds.
 \end{defn} 
 
In other words, the path   $ \path{S}{i,d}$
has the following property;  if a predicate expression $\ls{p}(\refvec(\ls{p}))$ along  $ \path{S}{i,d}$  is evaluated 
with respect to the predicate-free schema consisting of the 
sequence of assignments preceding  that predicate in $ \path{S}{i,d}$, then the 
value of  the resulting predicate term given by $i$ `agrees' with 
the value given in $ \path{S}{i,d}$.

By   Lemma~\ref{exe.lem}, this defines the path 
$\path{S}{i,d} \in \allpath(S)$ uniquely.

\begin{defn}[the semantics of arbitrary schemas] \label{genmeaning}
\rm
 If 
$\path{S}{i,d}$ is finite, we define 
$$\dd{S}{i}=\dd{\schema(\path{S}{i,d})}{i}$$ 
(which is already defined, since 
$ \schema(\path{S}{i,d})$ is predicate-free) otherwise 
$\path{S}{i,d}$ is infinite  and we define $\dd{S}{i}=\bot$. In this last case we may say that $\dd{S}{i}$ is not terminating.  Also,  for schemas $S,T$ and \ters\ $i$ and $j$  we write $\dd{S}{i}(\omega)=\dd{T}{j}(\omega)$ to mean $\dd{S}{i}=\bot \iff \dd{T}{j}=\bot$.  For convenience, if $S$ is predicate-free and $ d: \var \ra \term(\f,\var)$ is a state 
then we define unambiguously 
$ \dd{S}{}=\dd{S}{i}$; that is, we assume that the \ter\ $i$ is Herbrand if  $d$ is a Herbrand state; and we will write $\dd{\mu}{}$ to mean $\dd{\schema(\mu)}{}$ for any $\mu \in \lang(S)^*$. 
\end{defn} 

Observe that $\dd{S_1S_2}{i}={\mathcal M}\quine{S_2}^{i}_{\dd{S_1}{i}}$ and 
$$\path{S_1S_2}{i,d}=\path{S_1}{i,d}\path{S_2}{i,\dd{S_1}{i}}$$ hold for all schemas (not just predicate-free ones).

Given a  schema $S$,  let $\mu \in \pre(\pathset(S))$. We say that $\mu$ passes through a predicate term $p(\vt)$ 
if $\mu$ has a prefix $\mu'$ ending  in $\ul{\ls{p},Y}$ for $Y\in \tf$  \st\ $\mean{\mu'}{}{e}(\refvec(\ls{p})) = \vt$ holds. 
 We say that $p(\vt)= Y$ is  a \emph{consequence} of $\mu$ in this case. 

\section{Free and liberal schemas} \label{free.lib.sect}

Given an initial state and an \ter,\ a path through a schema defines a term $f(\vt)$ or a predicate term $p(\vt)$ at each symbol that it encounters. For this paper, we wish to consider the class of schemas for which no term or predicate term is defined more than once along any path, given $e$ as the initial state and assuming that all \ters\ are Herbrand. 

 \begin{defn}[free and liberal   schemas] \rm
 Let $ S$ be a schema.
\begin{itemize}
\item
  If for every  $\sigma\in \pre(\pathset(S))$ 
  there is a Herbrand \ter\ $i$ \st\ $\sigma \in \pre(\path{S}{i,e})$, 
   then 
$ S$ is said to be  \emph{free}.
\item 
If for every Herbrand \ter\ $i$ and any prefix 
$ \mu \; \ul{v \as \ls{f}(\va)}\;  \nu\;  \ul{w\as \ls[m]{g}(\vb)}\in \pre(\path{S}{i,e})$,   we have 
$$\ee{\mu\; \ul{v \as \ls{f}(\va)}}{}(v)\not= \ee{\mu\; \ul{v \as \ls{f}(\va)}\; \nu\; \ul{w\as \ls[m]{g}(\vb)}}{}(w),$$ then $ S$ is said to be  \emph{liberal}. (If $f\not=g$ then of course this  condition is trivially satisfied.)
\end{itemize}
\end{defn}

Thus a schema $S$  is said to be free if for every path through $S$, there is a Herbrand \ter\ which follows it with the natural state $e$ as the initial state, and a schema $S$ is said to be
 liberal if given any path through $S$ passing  through two \asss\ and a Herbrand \ter\  which follows it with  $e$ as the initial state,  the  \asss\ give distinct values to the variables to which they assign. The definitions of freeness and liberality were first given in \cite{paterson:thesis}. 
 
 Observe that if a schema $S$ is free, then given a Herbrand \ter\ $i$,
 $$\mu \; \ul{\ls{p},X} \; \mu'\; \ul{\ls[m]{p},Y} \; \in \pre(\path{S}{i,e})$$ implies that 
$$\ee{\mu}{}(\refvec(\ls{p})) \not= \ee{\mu\mu'}{}(\refvec(\ls[m]{p}))$$ holds, since otherwise there would be no Herbrand \ter\ whose path (for initial state $e$) has the  prefix 
 $\mu \; \ul{\ls{p},X}\; \mu' \; \ul{\ls[m]{p}, \neg X} $. Thus a path through a free schema cannot pass more than once (for initial state $e$)  through the same predicate term.  Hence if a Herbrand \ter\ $i$ maps only finitely many  predicate terms to $\tru$, and $S$ is a free schema,   then the path $\path{S}{i,e}$ terminates. Similarly,  if a schema $S$ is free and predicate-linear, and a Herbrand \ter\ $j$ maps finitely many \emph{while} predicate terms in $S$ to $\tru$, then  the path $\path{S}{j,e}$ terminates.


Proposition \ref{plonk.prop} demonstrates the use of requiring our schemas to be liberal. 

\begin{prop}\label{plonk.prop} 
Let $S,T_1,T_2$ be predicate-free schemas and assume that each schema $ST_i$ is liberal.  Let $v_1,v_2\in \var$. 
If $\ee{ST_1}{}(v_1) = \ee{ST_2}{}(v_2)$, then   $\ee{T_1}{}(v_1) = \ee{T_2}{}(v_2)$ holds.  
\end{prop}

\proof 
Assume $\ee{ST_1}{}(v_1) = \ee{ST_2}{}(v_2)$ holds. We will  prove $\ee{T_1}{}(v_1) = \ee{T_2}{}(v_2)$ by induction on the number of \asss\ in $T_1$. \Wma\ that each schema  $ST_i$ contains an \ass\ to $v_i$, since if this holds for exactly one value of  $i$, then a contradiction is obtained, and if it is false for both values of $i$, then the conclusion follows immediately.   
Write $$\ee{ST_1}{}(v_1) = \ee{ST_2}{}(v_2) = f(\vt)$$
and let $v_i \as f_i(\vu_i);$ be the last \ass\ to $v_i$ in $ST_i$ for each $i$.  Clearly $f_1=f_2=f$. 
\begin{itemize}
\item 
Suppose that  in the case of  $ST_1$, this last \ass\ to $v_1$  occurs in $S$. Thus this \ass\ sets the variable $v_1$ to $f(\vt)$.  Since $ST_2$ is liberal and $\ee{ST_2}{}(v_2) = f(\vt)$ holds, no \ass\ in $T_2$ can set a variable to $f(\vt)$ along $ST_2$, hence  $v_1 \as f(\vu_1);$  is also the last \ass\ to $v_1=v_2$ in $ST_2$, and so 
$\ee{T_1}{}(v_1) = \ee{T_2}{}(v_2)=v_1=v_2$ follows, thus proving the desired result. 
\item
Thus we may assume that the last \ass\ $v_1 \as f(\vu_1);$ to $v_1$ in $ST_1$ occurs in $T_1$. Similarly,  we may assume that the last \ass\ $v_2 \as f(\vu_2);$ to $v_2$ in $ST_2$ occurs in $T_2$.  Let $u_1$ and $u_2$ be the first components of $\vu_1$ and $\vu_2$ respectively, and write 
$T_i =T_i'\, v_i \as f_i(\vu_i); T_i''$ for each $i$. By the inductive hypothesis applied to $S$ and each  $T_i'$, the term
$\ee{T_i'}{}(u_i)$ is  the same  for each $i$; the Proposition then follows from the analogous result for the other components of each $\vu_i$. \proofend
\end{itemize}

Proposition  \ref{plonk.prop} need not hold for non-liberal schemas; for example, if $S$ and $T_1$ are both $v \as g();$ (so $ST_1$ is not liberal), $T_2= \ski$ and $v_1=v_2=v$.

As mentioned in the introduction, it was proved in \cite{paterson:thesis} that it is not  decidable whether an (unstructured)  schema is free, but it \emph{is} decidable whether it is   liberal, or liberal and free. Theorem \ref{lib.free.cond.thm} proves  the latter result for structured schemas.   It is an  open  question as to whether  freeness of a linear or function-linear  schema  is decidable.

\begin{thm}[it is decidable whether a schema is liberal and free] \label{lib.free.cond.thm} \mbox{} \\
Let $S$ be a schema. Then $S$ is both liberal and free \ifa\ for every path-segment $\tilde{x} \mu \tilde{y}$ in $S$  with $\tilde{x}, \tilde{y}\in \lang(S)$,  $\, \simbol(\tilde{x})=\simbol(\tilde{y})$ and  \st\  the same labelled symbol does not occur more than once in $\tilde{x} \mu$ or in $\mu \tilde{y}$, then 
either $\tilde{x}$ and $\tilde{y}$ reference a different vector of variables, or the path-segment $\tilde{x}\mu$ contains an \ass\ to a variable referenced by  $\tilde{y}$. 
\\
In particular, it is decidable whether a schema is both liberal and free.
\end{thm}

{\noindent{\it Proof }} \cite{paterson:thesis}. 
Assume that $S$ is both liberal and free. Then for any  path-segment $\tilde{x} \mu \tilde{y}$  satisfying the conditions given, there is a prefix $\Theta$ and a Herbrand \ter\ $i$  
 \st\ $\Theta \tilde{x} \mu \tilde{y} \in\pre(\path{S}{i,e})$, and distinct (predicate) terms are defined when $\tilde{x} $ and $ \tilde{y}$ are reached, thus proving the necessity of the condition. 
\\
 To prove sufficiency, first observe that the `non-repeating' condition on the letters of the path-segment $\mu$ may be ignored, since path-segments that begin and end with letters having the same labelled symbol can be removed from within $\tilde{x} \mu$ and $\mu \tilde{y}$ until it is satisfied. 
 Consider the set of prefixes of $\pathset(S)$ of the form $\Theta\tilde{x} \mu \tilde{y}$ with $\simbol(\tilde{x})=\simbol(\tilde{y})$ \st\ $\tilde{x} \mu \tilde{y}$ satisfies the condition given. By induction on the length of such prefixes, it can be shown that every \ass\ encountered along such a prefix defines a different term (for initial state $e$), and the result follows immediately  from this. 
\\
Since there are finitely many path-segments in $S$ satisfying the conditions given for $\tilde{x} \mu \tilde{y}$ and these can be enumerated,  the decidability of liberality and freeness for the set of schemas follows easily. 
 \proofend

 \section{Subschemas and Slicing Conditions} \label{equiv.defn.sect}

\begin{defn}[Subschemas  of a schema] \label{slice.defn} \rm
 The set of  subschemas   of a    schema $S$ is the minimal set of schemas  which satisfies    the following rules; 
\begin{itemize}
\item
$\ski$ is a subschema of any schema. 
\item
$S_1$ and $S_2$ are both subschemas of any schema $S_1 S_2$.
\item
If $S'$ is a subschema of $S$, then $S'T$ and $TS'$ are subschemas of $ST$ and $TS$ respectively. 
\item
if $T'$ is a subschema of $T$ then $\whi p(\vu) \du T'$ is a subschema of $\whi p(\vu) \du T$;
\item
 if $T'$ is a subschema of $T$ then  the \textit{if} schema  $\si q(\vu) \then S \els T'$ is a subschema of \\
 $\si q(\vu) \then S \els T$ (the true and false parts may be interchanged in this  example);
\item
 a subschema of a subschema of $S$ is itself a subschema of $S$. 
\end{itemize}
  \end{defn}

\begin{defn}[the semantic $u$-slice condition for  $u\in \var \cup \{\omega\}$] \rm
Let $T$ be a subschema of a schema $S$. Then given $u\in \var$, we say that 
$T$ is a $u$-slice of $S$ if  given any domain $D$, any state $ d:\var \ra D$  and any $ i \in \Int({\f}, {\p}, D)$, $\,\dd{S}{i} \not= \bot \Ra \,(\dd{T}{i} \not= \bot \wedge \, \dd{S}{i}(u)= \dd{T}{i}(u))$ holds.
  We also say that $T$ is an $\omega$-slice of $S$ if given any domain $D$, any state $ d:\var \ra D$  and any $ i \in \Int({\f}, {\p}, D)$,  $\,\dd{S}{i} \not= \bot \iff\,\dd{T}{i} \not= \bot$ holds. 
\end{defn}


Thus the  $u$-slice condition  is given  in terms of every conceivable domain and initial state; however it is well known that the Herbrand domain is the only one that needs to be considered when considering many schema problems. 
 Theorem \ref{freeint.thm}, which is virtually a restatement  of \cite[Theorem 4-1]{manna:book}, 
ensures that for slicing purposes,  we only need to consider Herbrand \ters\ and the natural state $e$.

\begin{thm}
\label{freeint.thm}
Let $\chi $ be a set of schemas, let $ D$ be a domain, let $d$ be a function from the set of variables into $D$ and let $i$ be an \ter\ using this domain.  Then there is a Herbrand \ter\ $j$ \st\ the following hold.
\begin{enumerate}
\item
 For all $S\in \chi$, the path $\path{S}{j,e}=\path{S}{i,d}$.
\item
If $S_1,S_2 \in \chi$ and $v_1,v_2$ are variables and  $\rho_k \in \pre(\path{S_k}{j,e})$ for $k=1,2$ and $\ee{\rho_1}{} (v_1)=\ee{\rho_2}{} (v_2)$, then also  $\dd{\rho_1}{i} (v_1)=\dd{\rho_2}{i} (v_2)$ holds. 
\end{enumerate}
\end{thm}

Throughout the remainder of the paper, all \ters\ will be assumed to be Herbrand.

 \section{The data dependence relations $\perc$ and $\percFinal$ and Weiser's labelled symbol set} \label{data.defn.sect}

Definition \ref{perc.defn} formalises the $\perc$ and $\percFinal$ relations introduced in Section \ref{relevance.slicing.subsect}.

 \begin{defn}[the  $\perc$ and $\percFinal$ relations and parameterised path-segments]\rm  \label{perc.defn}
Let $S$ be a   schema and  let $\sigma$ be a path-segment in $S$.
\\
 We call $\sigma$ an $F$-path-segment, or $vF$-path-segment for $F\in \f^*$ and $v\in \var$ if $\ee{\sigma}{}(u)$  for some $u\in \var$ is an $F$-term, or $vF$-term, respectively.  We also call these path-segments an $Fu$-path-segment or $vFu$-path-segment respectively.
\\
We call $\sigma \ul{\ls{p},Z}$ an $Fp$-path-segment or $F\ls{p}$-path-segment in $S$ if $\ee{\sigma}{}(u)$   is an $F$-term for some $u\in \var$ referenced by $\ls{p}$ in $S$. We define $vF\ls{p}$-path-segments analogously. 
\\
 We sometimes strengthen these definitions by  using \emph{labelled} function symbols in the word $F$ to indicate which labelled \ass\ in $S$  creates the appropriate subterm of $\ee{\sigma}{}(u)$. 
We write $\ls{f} \perc \ls[m]{g}$ if $S$ contains an $\ls{f} \ls[m]{g}$-path-segment for $f\in \f$ and $g\in \f \cup \p$, and write $\ls{f} \percFinal u$ if $S$ contains a terminal path-segment $\sigma$ \st\ $\ee{\sigma}{}(u)$ is an $f$-term.
\end{defn} 

The relations $\perc[]$ and $\percFinal[]$ correspond to  the data dependence relation in program slicing. 
 We now give examples of these relations.  If $S$ is the schema of Fig.
   \ref{not.lib.fig},  the path-segment $\ul{u\as f(u)}\; \ul{q(w),\tru} \;\ul{w \as h_1(w)} \; \ul{u\as h_2(u)}$ in $S$ is both an $fh_2$-path-segment and a $ufh_2$-path-segment, and the relation $f \perc h_2$ holds. Similarly, the path-segment $\ul{q(w),\tru} \;\ul{w \as h_1(w)} \; \ul{u\as h_2(u)}\;\ul{p(u),\tru}$ is a $uh_2p$-path-segment and an $h_2p$-path-segment, and $h_2 \perc p$ holds. Since 
$\ul{v\as g_1()}\;\ul{u\as f(u)}\; \ul{q(w),\fal}$ is a terminal path-segment in $S$, $g_1 \percFinal v$ holds.


  \begin{defn}[Weiser's labelled symbol set] \rm  \label{need.defn}
   Let  $S$ be a 
    schema and  let $u\in \{ \omega\} \cup \var $. Then we define 
   $\need(u)\subb \lfunc(S) \cup \lpred(S)$  to be the minimal set satisfying the  following conditions.
\begin{enumerate}
\item
If $\ls{f}\percFinal u\in \var$, then $\ls{f} \in \need(u)$ holds. 
\item
If $u =\omega$ then $\lwhipred(S) \subb \need(u)$. 
\item
If $x\in \need(u)$ and $\ls{f}\perc x$, then $\ls{f} \in \need(u)$ holds. 
\item 
If $x \in \need(u)$ and $\ls{p} \cont x$ then $\ls{p} \in \need(u)$. 
\end{enumerate}
   \end{defn} 
   
 The set $\need(u)$ (traditionally only defined for the case in which $u\in \var$, and for programs rather than schemas) is 
 fundamental to most slicing algorithms. It contains all symbols which might conceivably affect the final value of $u$ (if $u$ is a variable) or termination (if $u=\omega$). This assertion is formalised in Theorem \ref{weiser.thm}.

Given a schema $S$ and a set $\Sigma \subb \lsym[](S)$ satisfying $(x\in \Sigma\,\wedge \,\ls{p}\cont x) \Ra \ls{p}\in \Sigma$, there is a subschema $T$ of $S$ \st\ $\lsym[](T)=\Sigma$, obtained from $S$ by deleting all elements of $\lsym[](S)-\Sigma$ from $S$. This subschema is easily shown to be unique. In particular, 
for any $u\in \var \cup \{\omega\}$,  every schema $S$ has a unique subschema $T$ satisfying $\lsym[](T)=\need(u)$.   By Theorem \ref{lib.free.cond.thm}, if $S$ is both free and liberal, then so is $T$.

 \begin{thm}\label{weiser.thm}
  Let $S$ be any  schema, let  $u\in \var \cup\{\omega\}$ and let $T$ be a subschema of $S$. If $\lsym[](T) = \need(u)$,  then $T$ is a $u$-slice of $S$. 
 \end{thm}

\proof
Proved in \cite[Theorem 18]{laurence:flfl}.

If $S$ is liberal, free, and function-linear,  then a subschema $T$ of $S$  is the  $u$-slice of $S$ with the minimal number of labelled symbols \ifa\ $\lsym[](T) = \need(u)$ holds,  as was  proved in \cite{laurence:flfl}; but  in general this is false.   To see this, consider the schema $S$ in Figure \ref{identical.parts.fig}. It is clearly irrelevant whether $p(w)$ maps to $\tru$ or $\fal$, and hence the \ass\  $w\as h()$ may be deleted to give a $u$-slice.

\begin{figure}[t]
\begin{center}
$
\begin{array}{llll}
w\as h(); \\
\si  p(w) && \then \;\; u\as g();\\
&& \els 
 \;\;\; u\as g();
 \end{array}
 $ 
\end{center}
\caption{$h\in \need(u)$, but deleting the \ass\ $w\as h()$ gives a $u$-slice of $S$} \label{identical.parts.fig}
\end{figure}

Even if a schema is both free and linear, Weiser's algorithm need not give minimal slices. To see this, consider the  linear  schema $S$ of Figure \ref{not.lib.fig} which can easily be seen to be free.  Owing to the constant $g_1$-\ass,\ $S$ is not liberal; any path entering the true part  of $p$ more than once would assign the same value, $g_1()$, to  $v$ each time. 
\\
Since  $S$ contains the  $fh_2p$-path-segment $u\as f(u)\, \ul{q,\tru} \, w \as h_1(w)\, u \as h_2(u) \, \ul{p,\tru}$, and $p \cont g_1$ and $g_1 \percFinal v$ hold, 
 $f\in \need(v)$ follows; but  the subschema $S'$ of $S$ in which the \ass\ $u\as f(u);$ is deleted is a $v$-slice of $S$, since any \ter\ $j$ satisfying $\ee{S'}{j}(v) \not= \ee{S}{j}(v)$ would have to define a path $\path{S}{j,e}$ passing through the $f$-\ass\ (since otherwise the deletion of $f$ from $S$ would make no difference to $\ee{S}{j}(v)$), and so the value of $v$ would be thus fixed at $g_1()$.  

   \section{Couples of \ters}  \label{couple.section}

In order to establish which predicate symbols of a schema must be included in a slice in order to preserve our desired semantics, 
we define the notion of a $p$-couple for a predicate $p$.

    \begin{defn}[couples] \rm \label{couple.defn}
    Let $i,j$ be \ters\ and let $p\in \p$. We say that the set $\{i,j\}$ is a $p$-couple if there is a vector term $\vt$
     \st\  $i$ and $j$ differ only at the  predicate term 
    $p(\vt)$. 
 In this case we may also say that $\{i,j\}$ is a $p(\vt)$-couple. 
If a component of  $\vt$ is an $F$-term for $F\in \f^*$, then 
    $\{i,j\}$ is an $Fp$-couple. Given any $u\in \var $ and schema $S$, we  also say that $\{i,j\}$ is an $Fpu$-couple or $p(\vt) u$-couple for $S$ if also 
 $\ee{S}{i}(u) \not= \ee{S}{j}(u)$ and {both} sides terminate.   
Lastly, we may label $p$ (an $F\ls{p} u$-couple, or $\ls{p}(\vt) u$-couple for $S$) to indicate 
that  the paths $\path{S}{i,e} $ and $\path{S}{j,e}$ diverge at $\ls{p} $ (at which point the predicate  term  $p(\vt)$ is defined). 

 We also make analogous definitions if instead $u=\omega$; that is, if a set $\{i,j\}$ is a $p$-couple for a predicate symbol $p$, then  we say  $\{i,j\}$ is a $p\omega$-couple  for a schema $S$ if exactly one path in $\{\path{S}{i,e}, \path{S}{j,e}\}$ terminates. 
    \end{defn} 
    
      Note  that a $pu$-couple is simply an $Fpu$-couple with $F$ as the empty word. The existence of a $pu$-couple for a schema $S$ `witnesses' the fact that $p$ affects the semantics of $S$, as defined by $u$.  As an example of a $p$-couple, 
 let $i$ be an \ter\ that maps the predicate terms $q(w)$,  $q(h_1(w))$ and $p(h_2(u))$  to $\tru$, and maps $q(h_1(h_1(w)))$ and $p(h_2(f(h_2(u))))$ to $\fal$ and let the \ter\  $j$ be identical except that it maps $p(h_2(f(h_2(u))))$ to $\tru$. Then $\{i,j\}$ is a $p$-couple. If $S$ is the schema in Fig.  \ref{not.lib.fig}, then both paths  $\path{S}{i,e}, \path{S}{j,e}$ pass twice through the body of $q$, with  $\path{S}{i,e}$  passing through $g_1$ only on the first occasion, whereas $\path{S}{j,e}$  passes twice through $g_1$.  Since both \ters\ define the same final value for $v$, 
 $\{i,j\}$ is  not a $pv$-couple  for $S$. However, if $T$ is the schema obtained from $S$  by replacing the \ass\  $v \as g_1();$ by $v\as g_2(v)$, then $\{i,j\}$ is   a $pv$-couple  for $T$.

Proposition \ref{slice.keep.couple.prop} follows immediately from Definition \ref{couple.defn}. 

\begin{prop} \label{slice.keep.couple.prop}
If   $u\in \var$ and $T$ is a $u$-slice of a schema $S$, then 
 a $pu$-couple for $S$ is also a $pu$-couple for $T$. \proofend
    \end{prop}

\begin{defn}[head and tails  of a couple]\rm  \label{head.tails.defn}
Let $S$ be a schema.
Let $u\in \var $,  and let $q\in \pred(S)$. Let $I=\{i,j\}$ be a $qu$-couple for $S$ and write $$\path{S}{k,e}=\mu \ul{q\labl,Z_k} \;\rho_k$$ for each $k\in I$ and $\{Z_i,Z_j\}=\tf$;  then we define 
$\tail(I,k)=\rho_k$ for each $k\in I$, and  $\mu=\head(I)$. 
\end{defn}

The motivation for Definition \ref{head.tails.defn} is given by Lemma \ref{change.prefix.lem}, which shows that given a  $pu$-couple for a free liberal  schema, a new $pu$-couple  may be obtained from it by replacing its head  by any prefix leading to  $p$, while keeping the same tails.

\begin{lem}[Changing the head of a couple]\label{change.prefix.lem}
Let $S$ be a free liberal   schema and let $p\labl\in \lpred(S)$ and 
 $u \in \var $. Suppose there is a $p\labl u$-couple $I$ for $S$ 
 and a prefix $\mu \, \ul{p\labl,\tru }$ in $S$, then there is a $pu$-couple $I'$ for $S$ \st\  
 $\mu=\head(I')$ and $ \{\tail(I,k)\vert\,k\in I\} = \{\tail(I',k)\vert\,k\in I'\}$. In particular, if there is a $p\labl u$-couple $I$ for $S$ and  $S$ contains an $Fp\labl$-path-segment for $F\in \f^*$, then \txs\ an $Fp\labl u$-couple $I'$ for $S$. 
\end{lem}

\proof
Write $I=\{i,j\}$. 
Since $S$ is free, \tx\ \ters\ $i',j'$ defining paths $\mu \ul{p\labl, Z}\tail(I,i)$
 and $\mu \ul{p\labl, \neg Z}\tail(I,j)$ for $Z\in \tf$, and by Proposition \ref{plonk.prop}, the final  value of $u$ after each path is still distinct. Thus it suffices to prove that $i',j'$ need not differ on any predicate term except the $p$-predicate term defined after $\mu$. However, if this is false, then $q(\vt')=Y$ must be a consequence of one of the paths and $q(\vt')=\neg Y$ must be a consequence of the other, for some predicate term $q(\vt')$ and $Y\in \tf$. Again, since $S$ is free, $q(\vt')$ must occur on the tails of both paths, and by Proposition \ref{plonk.prop} applied to the variables referenced by the appropriate occurrences of $q$ on each path and the prefixes of the paths preceding these occurrences, the same incompatibility would contradict the existence of the $p\labl u$-couple $I$. Thus we may define $I'=\{i',j'\}$. \proofend

For the remainder of this paper,  we use the following terminology with \ters.\ If $i$ is an \ter,\ $p(\vt)$ is a predicate term and $X\in \tf$,   then $i(p(\vt)=X)$ is the \ter\ which maps every predicate term to the same value as $i$ except $p(\vt)$, which it maps to $X$.

 Lemma \ref{change.prefix.lem} need not hold for schemas that are not both free and liberal.  To see this, consider the free, linear, non-liberal schema $S$ of Figure \ref{not.lib.fig}. 
\\
Let the \ter\ $i$ satisfy $q^i(t)=\tru$ \ifa\ the term $t=w$, and $p^i(h_2(u))=\tru$. If the \ter\ $j=i(p(h_2(u))=\fal)$, then $\set{i,j}$ is 
an $h_2pv$-couple for $S$, since $\ee{S}{i}(v)=g_1()$ whereas $\ee{S}{j}(v)=v$, 
 but there is no $fh_2pv$-couple for $S$, although $S$ contains an $fh_2p$-path-segment, since any \ter\ $k$ \st\ $\path{S}{k,e}$ passes through the $f$-\ass\  must satisfy $\ee{S}{k}(v)=g_1()$.

\section{Restriction to Special Schemas} \label{special.sect}

In order to prove our main results, we need to exclude from consideration schemas such as the one in Figure \ref{identical.parts.fig}. Therefore we will now only consider schemas \st\ if the same function symbol occurs in both parts of any if predicate, then the occurences assign to different variables. The utility of this assumption is demonstrated by Proposition \ref{opp.parts.prop}.

\begin{defn}[Special schemas] \rm 
Let $S$ be a predicate-linear free liberal schema. We say that $S$ is \emph{special} if given any $p\in \ifpred(S)$ and $f\in \f$ \st\ $p \cont \ls{f}\, (\tru)$ and $p \cont \ls[m]{f}\, (\fal)$ hold, $\, \assig(\ls{f}) \not= \assig(\ls[m]{f})$ holds. 
\end{defn}

Figure \ref{omega.not.min.fig} in Section \ref{omega.sect} gives an example of a special  schema.

\begin{prop} \label{opp.parts.prop}
Let $v\in \var$ and let $R,\, S_1,S_2$ be predicate-free schemas \st\ either $S_1$ or $S_2$ contains an \ass\ to $v$,  each schema $R S_j$ is liberal and for all $f\in \f$, if $S_1$ and $S_2$ both contain  \asss\ with  function symbol $f$, then they assign to different variables. Then $\ee{R S_1}{}(v)\not=\ee{R S_2}{}(v)$ holds. 
\end{prop}

\proof
If only one schema in the set $\{S_1,S_2\}$ contains an \ass\ to $v$, then the result follows from the liberality condition. If both do, let $f_j$ be the function symbol of the last \ass\ to $v$ in each  $S_j$. By our hypotheses, $f_1\not= f_2$, and each term $\ee{R S_j}{}(v)$ has $f_j$ as the outermost function symbol, giving the result. 
\proofend

\section{Main Theorems} \label{main.sect}

We wish to prove that for any $u\in \var$,  every schema which is a $u$-slice of  a given special schema $S$ contains every  symbol occurring in  $\need(u)$. Thus we need to refer to the recursive definition of $\need(u)$.  This motivates Lemmas \ref{pred.abuv.pred.lem},  \ref{pred.to.pred.lem} and \ref{pred.to.var.lem}, and Definition \ref{link.defn}.  We first consider Condition (4) in Definition \ref{need.defn}, and show that the property of defining a $pu$-couple is `backward-preserved' by the $\cont$ relation.


\begin{lem}\label{pred.abuv.pred.lem}
Let $S$ be a free  predicate-linear schema and assume $p \cont q$ for $p,q \in \pred(S)$. Let $u\in \var$. Assume that \txs\ a $qu$-couple for $S$. Then \txs\ a $pu$-couple for $S$. 
\end{lem}

\proof
Assume $p\cont q\, (X)$ holds and for each $r \in \pred(S)$, choose $Z_r \in \tf$, subject to the provisos that $Z_p =X$ and $r \in \whipred(S) \, \Rightarrow Z_r= \tru$. Since $X= \tru$ if $p$ is a \textit{while} predicate, this is  possible.

Since \txs\ a $qu$-couple for $S$, a pair of \ters\ $ (i,j)$ can be chosen \st\ $\{i,j\}$ is a $qu$-couple for $S$
and  the number of  predicate terms $r(\vs)$ that $i$  maps to $Z_r$ is minimal for all such pairs; clearly this number is  finite, since  the path $\path{S}{i,e}$ terminates. The path $\path{S}{i,e}$  must  pass through $q$ and hence through $\ul{p ,X}$, since $p\cont q\, (X)$ holds,  and hence there is a  predicate term $p( \vt )$ which $i$ maps to $X$. Since $q \not= p$, $p^j( \vt )=X$ also holds.  Define the \ters\ $i',j'$ to be identical to $i$ and $j$ respectively except that $i',j'$ both map $p( \vt )$ to $\neg X$. Thus $i'$  maps fewer predicate terms $r(\vs)$ to  $Z_r$ than $i$ does, and hence  by the minimality assumption on $i$, $\{i',j'\}$ is not a $qu$-couple for $S$. Hence either $$\ee{S}{i}(u)\not= \ee{S}{i'}(u) \text{ or }\ee{S}{j}(u)\not= \ee{S}{j'}(u)$$ holds. 

By the freeness of $S$ and the fact that $i'$ and $j'$ map finitely many predicate terms $r(\vs)$ for $r\in \whipred(S)$ to $\tru$, the paths $\path{S}{i',e}$ and $\path{S}{j',e}$ are both terminating, and so either $ \lbrace i,i' \rbrace$ or $ \lbrace j,j' \rbrace $ is a $pu$-couple for $S$,
 giving the result.
\proofend

It is convenient to make the following definitions, which merely give an alternative way of expressing Weiser's set.

\begin{defn}[$(p,X)$-links and $v$-feeding path-segments]\rm \label{link.defn}
Let $S$ be a predicate-linear schema. 
\\
Let $p\in \ifpred(S)$ and $X\in \tf$. A $(p,X)$-link in $S$ is a path-segment $\ul{p,X} \nu$ for some path  $\nu$ in the $X$-part of $p$ in $S$. 
\\
If $p\in \whipred(S)$, then the path-segment  $\ul{p,\fal }$ is called a $(p,\fal)$-link in $S$;   and  a path-segment in  $(\ul{p, \tru} \pathset(\body(p)))^* \ul{p,\fal }$ which passes at least once through $\pathset(\body(p))$ is a $(p,\tru)$-link. 
\\
Let $p,q\in \pred(S)$ and let $v\in \var$.  We say that a path-segment $\mu$ in $S$ $v$-feeds $p$ to $q$ if \txs\ $X\in \tf$ \st\  $\nu \mu \ul{q,\tru}$ is a path-segment in $S$ for some $(p,X)$-link $\nu$  and $\ee{\mu}{}(w)$ is a $vF$-term for some $F\in \f^*$ and $q$ references the variable $w$.  
\end{defn}

\begin{prop} \label{change.term.prop}
Let $S_1,S_2,T$ be predicate-free schemas and let $v,w$ be variables \st\ $\ee{S_1}{}(v)\not= \ee{S_2}{}(v)$ and assume that $\ee{T}{}(w)$ is a $vG$-term for some $G\in \f^*$. Then $\ee{S_1T}{}(w)\not= \ee{S_2T}{}(w)$ holds. 
\end{prop}

\proof
This follows by induction on the total number of \asss\ and occurrences of $\ski$ in  $T$. If  $T=\ski$ then $v=vG=w$ and the result is straightforward. If  $T=T'\ski$ or $T=T'\, w'\as g(\vu);$ for $w'\not= w$, then 
$\ee{S_iT}{}(w)=\ee{S_iT'}{}(w)$ for each $i$ and so  the result follows from the inductive hypothesis applied to $T'$. Thus \wma\ that 
 $T=T'\, w\as g(w_1,\ldots,w_m);$. Hence we may write $G=G'g$ \st\  for  some $j\le m$, $\,\ee{T'}{}(w_j)$ is a $vG'$-term. From the inductive hypothesis applied to $T'$, $\ee{S_1T'}{}(w_j)\not= \ee{S_2T'}{}(w_j)$ holds. Since $\ee{S_iT}{}(w)= g(\ee{S_iT'}{}(w_1),\ldots, \ee{S_iT'}{}(w_m))$ for each $i$, the result follows. 
\proofend

We can now prove that the property of defining a $pu$-couple is `backward-preserved' by the transitive closure of Conditions (3) and (4) of  Definition \ref{need.defn}.

\begin{lem}\label{pred.to.pred.lem}
Let $S$ be a special schema. Let $u,v \in \var $ and $p,q\in \pred(S)$. Assume that \txs\  a  $ qu$-couple for $S$.   Suppose that \txs\  an \ass\ to  $v$ in the body or in one part of $p$ in $S$ and that \txs\ a path-segment in $S$ 
$v$-feeding $p$ to $q$. Then 
  \txs\ a $pu$-couple for $S$.
\end{lem}

\proof   Given a fixed pair $(p,u)$, 
we will assume that the conclusion of the Lemma is false, but that  the hypotheses are true for some triple $(q,v,\sigma)$, where  $\sigma$ is a path-segment in $S$
$v$-feeding $p$ to $q$, and will  show that this leads to a contradiction. 
 We will assume that the triple $(q,v,\sigma)$ is chosen  \st\ the path-segment $\sigma$ is of minimal length   \st\ the hypotheses of the Lemma  are satisfied.
\\
For some $X\in \tf$, let $\rho$ be a $(p,X)$-link passing through an \ass\ to $v$ and   let $\mu \rho \sigma\in \pre(\pathset(S))$. 
By Lemma \ref{change.prefix.lem}, we can choose a $ qu$-couple $I = \{i,j\}$ for $S$ \st\ $\head(I)=\mu \rho \sigma$.
We may assume that  $i$ and $j$ map finitely many \textit{while} predicate terms to $\tru$, since the \ters\ define terminating paths. 
 Let  $m$ be the total number of $r$-predicate terms   which $i$ and $j$ both map to $\tru$, 
 where $r$ is  the \textit{while} predicate lying immediately above $q$ if $q\in \ifpred(S)$, or $q$ itself if $q\in \whipred(S)$. If $q\in \ifpred(S)$ and $q$ does not lie in the body of a \textit{while} predicate, then $m$ and $r$ are undefined. 
We assume  that $I$ is chosen \st\ if defined,  $m$ is minimal for the chosen values of $q$, $v$ and $\sigma$. 

 Let $\rho'$ be any $(p,\neg X)$-link and let $\Gamma$ be the set of all pairs $(\tilde{q}(\vtt),Z)$ \st\  $\tilde{q}(\vtt)=Z$ is a consequence of the prefix $\mu \rho' \sigma$, but is not  a consequence of  $\mu \rho \sigma$, and
let  the \ters\ $i',j'$ be obtained  by altering $i$ and $j$ respectively in accordance with the pairs in $\Gamma$; thus, if $(\tilde{q}(\vtt),Z)\in \Gamma$ then ${\tilde{q}}^{i'}(\vtt)=Z$, otherwise ${\tilde{q}}^{i'}(\vtt)={\tilde{q}}^{i}(\vtt)$, and similarly for $j'$. Thus the paths  $\path{S}{i',e}$ and $\path{S}{j',e}$ both   have $\mu \rho' \sigma$ as a prefix. By the freeness of $S$, the set $\Gamma$ does not contain any subset of the form  $\{(\tilde{q}(\vtt),Z), (\tilde{q}(\vtt),\neg Z)\}$ and so $i'$ and $j'$ are well-defined. We write $I'=\{i',j'\}$.   
 We now show that a contradiction is obtained. The proof proceeds in stages. 
\begin{enumerate}
\item
For any   $(\tilde{q}(\vtt),Z)\in \Gamma$, we now show that there is no $\tilde{q}u$-couple for $S$. Assume this is false for some $(\tilde{q}(\vtt),Z)$.   By the definition of $\Gamma$, $\,\tilde{q}(\vtt)$ does not occur on $\mu$, and by  Lemma \ref{pred.abuv.pred.lem} and the fact that $p\not= \tilde q$ by the falsity of the conclusion of the Lemma,  $\tilde{q}(\vtt)$ does not occur on $\mu\rho'$ either,  and so    $\mu \rho' \sigma$ has a prefix $\mu \rho' \sigma'\ul{\tilde{q},Z}$  \st\ $\tilde{q}$ defines $\tilde{q}(\vtt)$   after $\mu \rho' \sigma'$ and since $\tilde{q}(\vtt)=Z$ is  not  a consequence of  $\mu \rho \sigma$, replacing $\rho$ by $\rho'$ in $\mu \rho \sigma'$ changes the $\tilde{q}$-predicate term defined after $\mu \rho \sigma'$. Hence for some variable $v'$ in the body or in one part of $p$, $\sigma'$ $v'$-feeds $p$ to $\tilde{q}$, contradicting the minimality of $\sigma$. 
\item
We now show that $I'$ is   a $qu$-couple for $S$.
Suppose  this is false.  Since $I$ is   a $qu$-couple  for $S$,  either $\ee{S}{i}(u)\not=\ee{S}{i'}(u)$ or the analogous assertion holds for $j$ and $j'$. However, since $S$ is free, changing $i$ or $j$ at finitely many predicate terms still results in an \ter\ defining a terminating path through $S$, and by 
 (1), does not change the final value of $u$ if the predicate terms have the form $\tilde{q}(\hat{\vt})$ for some $(\tilde{q}(\vtt),Z)\in \Gamma$, thus contradicting   the definitions of $i'$ and $j'$ immediately. 
\item
Hence $I'$ is a $qu$-couple for $S$. Let  $\vt= \ee{\mu \rho \sigma}{}(\refvec(q))$; thus $i$ and $j$ differ only at $q(\vt)$.  Clearly 
$i'$ and $ j'$ also  differ only  at $q(\vt)$ and so their paths diverge at $q(\vt)$. Since $S$ is free, $q(\vt)=Z$ is  not
a consequence of $\mu \rho \sigma$ for either $Z$, and so by (1) and the definition of $\Gamma$, $q(\vt)$ does not occur on $\mu \rho' \sigma$ either. Also, $\ee{\mu\rho\sigma}{}(w)\not=\ee{\mu\rho'\sigma}{}(w)$ holds for at least one variable $w$ referenced by $q$,  by the assumptions on $\rho$ and $\sigma$ and Proposition \ref{opp.parts.prop} applied to $\schema(\mu)$, $\schema(\rho)$ and $\schema(\rho')$,  and Proposition  \ref{change.term.prop} 
applied to $\schema(\mu\rho)$, $\schema(\mu\rho')$ and $\schema(\sigma)$, and so $q$ does not define  $q(\vt)$ after $\mu\rho'\sigma$. 
Thus $\path{S}{i',e}$ and  $\path{S}{j',e}$ pass at least twice through $q$ after $\mu \rho' \sigma$, and $m$ and $r$ are defined and 
$\head(I')= \mu \rho' \sigma \tau$ for some path-segment $\tau$ passing at least once through $\ul{r,\tru}$. 
\item
Thus by Lemma \ref{change.prefix.lem}, \txs\ a $qu$-couple $\tilde I=\{\tilde i,\tilde j\}$ for $S$ which has the same pair of tails as $I'$ and  \st\ $\head(\tilde I)=\mu \rho' \sigma$. \Wma\ that each $r$-predicate term which is not a consequence of either path   $\path{S}{\tilde i,e}$ or  $\path{S}{\tilde j,e}$  is mapped   to $\fal$ by both \ters\ in  $\tilde I$. We now show that this `cutting out' of the path-segment $\tau$ passing  through $\ul{r,\tru}$ from  $\head(I')$ contradicts the minimality of $m$. By (1) and   Lemma \ref{pred.abuv.pred.lem}, the elements of  $I'$ map the same number of $r$-predicate terms to $\tru$  as those in  $I$ do. Thus it suffices to prove that the \ters\ in $\tilde I$ map fewer $r$-predicate terms to $\tru$ than those in $I'$. 
By the freeness of $S$ and our assumption on $\tilde I$, the number of $r$-predicate terms mapped to $\tru$ by both  \ters\ in  $\tilde I$ is obtained by adding up the number of occurrences of $\ul{r,\tru}$ on $\head(\tilde I)$ to those on either tail of $\tilde I$, and subtracting the number of $r$-predicate terms mapped to $\tru$ occurring on both tails of $\tilde I$. The analogous assertion holds for $I'$. 
Clearly  $\head(\tilde I)$ has fewer occurrences of  $\ul{r,\tru}$ than  $\head(I')$ has. Since $I'$ and $\tilde I$ have the same tails, it thus  remains only to prove that the same number of $r$-predicate terms mapping to $\tru$ occur on both $\tail(I',i')$ and  $\tail(I',j')$ after $\head(I')$ as after $\head(\tilde I)$, and this follows from Proposition \ref{plonk.prop}, since replacing the prefix $\head(I')$ by $\head(\tilde I)$ preserves equalities between predicate terms occurring along $\tail(I',i')$ and  $\tail(I',j')$. 
\proofend 
\end{enumerate}

We now use Lemma \ref{pred.to.pred.lem} to prove the existence of a $pu$-couple where membership of the predicate $p$ in $\need(u)$ is witnessed by iteration of Conditions (1) and (4) of  Definition \ref{need.defn}.

\begin{lem} \label{pred.to.var.lem}
Let $S$ be a special schema. Let $u, v\in \var $ and $p\in \pred(S)$. Suppose that \txs\  an \ass\ to  $v$ in the body or in one part of $p$ in $S$ and  that \txs\ a terminal path-segment $\sigma $ in $S$ \st\  for some $G\in \f^*$, $\ee{\sigma}{}(u)$ is a $vG$-term. 
 Then 
  \txs\ a $pu$-couple for $S$.
\end{lem}

\proof
Let $T$ be the schema $S \si q(u) \then\ u \as g_1(); \els u\as g_2();$, where $q,g_1,g_2$ are distinct symbols not occurring in $S$. Clearly $T$ is special and the path-segment $\sigma$ $v$-feeds $p$ to $q$ in $T$. 
The result follows from Lemma \ref{pred.to.pred.lem} applied to $T$. 
\proofend

 We now use the preceding two Lemmas to show that every symbol  of $\need(u)$ for a special schema $S$ can affect the semantics of $S$.

\begin{thm}\label{need.thm}
Let $S$ be a special schema. Let $ u\in \var $. 
\begin{enumerate}
\item
For all $p\in \need(u)\cap \p$ \txs\ a $pu$-couple for $S$. 
\item
 For all $f\labl\in \need(u) \cap \f\labL$, either \txs\ an \ter\ $i$ \st\ the term  $\ee{S}{i}(u)$ contains the symbol $f$, or \txs\ $p\in \need(u) \cap \p$ \st\ \txs\ a $p(\vt)u$-couple for $S$ for some vector term $\vt$ containing $f$. 
\end{enumerate}
\end{thm}

\proof

Let $\Theta$ be the set of all predicates $p$ in $S$ \st\ \txs\ a $pu$-couple for $S$
  and let  $P=\need(u) \cap \p$.
\begin{enumerate}
\item
 Observe that from Conditions (1,3,4) of Definition \ref{need.defn}, $P$ is the minimal subset of $\pred(S)$ satisfying the following two conditions.
\begin{itemize}
\item
 If $p\in \pred(S)$ and $ p\cont \ls{f}$ for a labelled function symbol $\ls{f}$ and \txs\  a terminal $\ls{f}F u$-path-segment for some $F\in \f\labL^*$, then $p\in P$ holds. 
\item
If $p\in \pred(S)$ and $ p\cont \ls{f}$ for a labelled function symbol $\ls{f}$ and 
$ q\in P$ and $S$ contains an $\ls{f}F q$-path-segment for some $F\in \f\labL^*$, then $p\in P$.
\end{itemize}
 By Lemmas \ref{pred.to.var.lem} and \ref{pred.to.pred.lem} respectively,  $\Theta$ also satisfies both these conditions; hence $P \subb \Theta$, as required. 
\item
If $\ls{f}\in \need(u)\cap \f\labL$, then from Definition \ref{need.defn}, one of the following two possibilities must occur. 
\begin{itemize}
\item
 \Txs\ an $\ls{f}F u$-path-segment for some 
$F\in \f\labL^*$, in which case by the freeness of $S$ \txs\ an \ter\ $i$ \st\ the term  $\ee{S}{i}(u)$ contains the symbol $f$, as required. 
\item
 The schema $S$ contains an $\ls{f}F p$-path-segment for some $F\in \f\labL^*$ and $p\in P\subb \Theta$ holds by  Part (1) of this Theorem, in which case by Lemma \ref{change.prefix.lem}, \txs\ a  $p(\vt)u$-couple for $S$ for some vector term $\vt$ one of whose components is an $fF$-term, proving the result.  
\end{itemize}
\end{enumerate}

\proofend

The main  theorem of the paper follows. 

\begin{thm}\label{main1.thm}
Let $S$ be a special schema. Let $ u\in \var $ and  let $T$ be a subschema of $S$. 
\begin{enumerate}
\item
If $\lsym[](T)= \need(u)$ then $T$ is a $u$-slice of $S$.
\item
If $T$ is a $u$-slice of $S$, then  $T$ contains at least one occurrence of every symbol in $\need(u)$. In particular, if $\lsym[](T)= \need(u)$, then no  subschema $T'$ of $T$ satisfying $T'\not=T$ is a $u$-slice of $S$ unless \txs\ $f\in \func(T)$ \st\  $T$ contains at least two occurrences of $f$ and $T'$ contains at least one, but not all occurrences of $f$ lying in $T$.
\end{enumerate}
\end{thm}

\proof 
Part (1) is a restatement of Theorem \ref{weiser.thm} for the subclass of special schemas. 
Part (2) follows immediately from
 Theorem \ref{need.thm} and Proposition  \ref{slice.keep.couple.prop}, and the definition of a $u$-slice. 
\proofend

\section{Weiser's algorithm does not give minimal $\omega$-slices for Special Schemas} \label{omega.sect}

\begin{figure}[h]

\begin{center}
$ 
\begin{array}{llll}
\; x\as c(); \\
\si p(x)  \then &\{ \\
&u\as g_1();\\
&v\as g_2();\\
&\} \\
\;\;\;\;\;\;\;\;\;\;\els &\{ \\
&v\as g_1();\\
&u\as g_2();\\
&\} \\
\;w\as f(u); \\
\whi q(w)  \du  & \{ \\
 & w \as f(v);  \\
& a \as h(a);  \\
& v \as k(a);  \\
&\}
 \end{array}
 $
\end{center}
\caption{Deleting the \ass\ $ x\as c();$  gives an $\omega$-slice of this special schema, although $c\in \need(\omega)$} \label{omega.not.min.fig}
\end{figure}

Theorems \ref{need.thm} and Part (2) of Theorem \ref{main1.thm} do not hold if the variable $u$ is replaced by $\omega$. To see this, consider the special schema $S$ of Figure \ref{omega.not.min.fig}. By iterating Conditions (2,3,4) of  Definition \ref{need.defn}, it follows that  $\need(\omega)$ contains both occurrences of each of $f,g_1,g_2$ and hence also contains $p$ and $c$, 
 but we now show that there is no $p\omega$-couple for $S$. For suppose that $\{i,j\}$ is a $p\omega$-couple for $S$, and so $i$ and $j$ define paths passing different ways through $p$. Let $\Omega =\{\path{S}{i,e},\,\path{S}{j,e}\}$.  Observe that one path in $\Omega$  defines the same predicate term on the second occasion that it passes through $q$ as the other does on the first occasion, and that  if $n \ge 3$, the two paths in $\Omega$ define the same  predicate term on the $n$th occasion that they pass through $q$. Thus suppose that one path terminates after passing $m$ times through $q$. If  $m\in \{1,2\}$, then  the other also terminates after passing not more than $3-m$ times through $q$. If $m \ge 3$, then so does the other after passing  not more than $m$ times through $q$, giving a contradiction. 
\\
Thus Part (1) of Theorem \ref{need.thm} is false in this case, and hence it follows easily that 
 the subschema of $S$ obtained by deleting the  \ass\ $ x\as c();$ is an $\omega$-slice.

\section{Conclusions and suggestions for further work}
\label{conclu.sect}

We have shown that for any variable $u$ and a special schema $S$, the    subschema  $T$ of $S$ containing the set of predicate symbols and labelled function symbols in the `Weiser set' $\need(u)$, and no others, has the minimal set of predicate and function symbols of any $u$-slice of $S$.  

This leaves open the possibility that there exists a  subschema of $T$ that is a $u$-slice of $S$ and has fewer, but still non-zero, occurrences of some of the function symbols occurring with labels in $\need(u)$.   It is not clear whether an example of a special schema exists with this property. Further research should investigate this problem. However if $S$ is not special, this can certainly happen, as the example of the predicate-linear but non-liberal schema in Figure \ref{not.linear.fig} shows.

For $u=\omega$, we have shown that the corresponding result fails, as the special  schema in Figure \ref{omega.not.min.fig} shows. The existence of this special schema does, however, 
  show the strengthening of our main result compared to that of \cite{sdetal:lpr}. 
Further work will also  concentrate on obtaining minimal $u$-slices for  larger classes of schemas. In particular, it would be of interest to be able to effectively characterise minimal slices for a reasonable class of schemas containing those in Figures   \ref{not.linear.fig} and \ref{not.lib.fig}, which are near-liberal but not liberal. \cite{laurence:danicic:hierons:nlfl:jlap} gives a related decidability result for schema equivalence. 
In addition, the main theorem of the paper can almost certainly be generalised to allow slicing criteria according to which the value of a given variable at a particular point within a program must be preserved by a slice, rather than at the end.

{ \bf Acknowledgements}

This work was supported by a grant from the Engineering and Physical Sciences Research Council, Grant EP/E002919/1.

\bibliographystyle{elsart-num}
\bibliography{slice240605}

\end{document}